\documentstyle{book}
\newcommand{\hide}[1]{}
\begin{document}

\markboth{{\bf Indian Contributions To HEP In The 20$^{th}$
Century}} {{G. Rajasekaran}}

\begin{center}
{\large{\bf Indian Contributions to High Energy Physics in the 20th Century}}
\vskip0.5cm
{\bf G. Rajasekaran}
\vskip0.35cm
{\it Institute of Mathematical Sciences, Madras 600113.\\
e-mail: graj@imsc.ernet.in}
\vskip0.35cm
\end{center}

\vspace{1cm}
\underline{Abstract} :The inward bound path of discovery unravelling the
mysteries of matter and the forces that hold it together has culminated
at the end of the twentieth century in a theory of the Fundamental Forces
of Nature based on Nonabelian Gauge Fields, called the Standard Model of
High Energy Physics. In this article we trace the historical development
of the ideas and the experimental discoveries on which this theory is 
based. We also mark significant Indian contributions wherever possible.
Finally we have a glimpse at future developments. An Appendix on more
Indian contributions is added at the end.

\vspace{1cm}
\underline{A disclaimer} :  This is  really a job for a
historian of science.  The selection of this topic was not done by me. The
topic and the title were given to me.  I have reinterpreted the title.
I am not interpreting it in a narrow sense. That would be suicidal and a sure
prescription for turning everybody into my enemy. I have chosen a broad
canvas and a broad brush. So, your portrait  may be too small to show here.
In any case, this is not a selection committee and I am not assessing anybody
for recruitment or promotion. All that I will try to do is to give my view of
the development of Fundamental Physics in the 20th century and, on the way,
refer to India's contributions (if any).\\

\vspace{1cm}
\underline {Plan of the talk}

\begin{enumerate}
\item  Scope
\item HEP before and after the Standard Model
\item Future of HEP
\item Present Status in India and Suggestions for Future

\item A  list of more Indian contributions

\end{enumerate}

\vspace{1cm}

(This was an invited talk at the XIV DAE Symposium on High Energy Physics, 
University of Hyderabad,
Dec 2000, published in the Proceedings (Eds:A K Kapoor, P K Panigrahi and
Bindu A Bambah) p1.)

\newpage

\section{Scope}

The earlier part of the 20th Century was marked by two revolutions that
rocked the Foundations of Physics.

\begin{center}

\begin{tabular}{|c|} \hline \\

1. \ Quantum Mechanics  \ \& \ 2. \ Relativity \\

\\ \hline

\end{tabular}

\end{center}

\noindent Quantum Mechanics became the basis for understanding Atoms,
and then, coupled with Special Relativity, Quantum Mechanics provided
the framework for understanding the Atomic Nucleus and what lies inside.

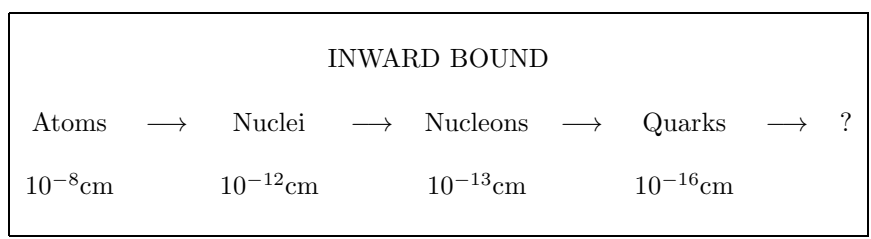
\begin{center}
\begin{tabular}{|ccccccccc|} \hline
& & & & & & & & \\
\multicolumn{9}{|c|}{INWARD BOUND} \\
& & & & & & & & \\
Atoms & $\longrightarrow$ & Nuclei & $\longrightarrow$ & Nucleons &
$\longrightarrow$ & Quarks & $\longrightarrow$ & ? \\
& & & & & & & & \\
10$^{-8}$cm &  & 10$^{-12}$cm & &  10$^{-13}$cm & &  10$^{-16}$cm &  &
\\
& & & & & & & & \\ \hline
\end{tabular}
\end{center}

\noindent
This inward bound path of discovery unraveling the mysteries of matter
and the forces holding it together -- at deeper and ever deeper levels
-- has culminated, at the end of the 20th century, in the theory of {\it
Fundamental Forces} based on {\it Nonabelian Gauge Fields}, for which we
have given a rather prosaic name :

\begin{center}
\begin{tabular}{|c|} \hline \\
THE STANDARD MODEL OF HIGH ENERGY PHYSICS \\
\\ \hline
\end{tabular}
\end{center}

\noindent
But, this is not the end of the road. More on that, later.
Thus, HEP is just the continuation of the era of discoveries that saw
the discovery of the {\it electron}, the discovery of {\it
radioactivity} and {\it X rays}, the discovery of the {\it nucleus} and
the {\it neutron} and the discovery of {\it cosmic rays} and the {\it
positron}.\\

These discoveries went hand in hand with the development of {\it Quantum
Mechanics, Relativity} and {\it Quantum Field Theory}. For, without the
conceptual advances made in these theoretical developments, the above
experimental discoveries could not have been assimilated into the {\it
framework of Physics}.\\

So, the present-day HEP must be regarded as the successor to Nuclear
Physics, which in turn was the successor to Atomic Physics :
\begin{center}

\begin{tabular}{|c|} \hline \\
Atomic Physics $\longrightarrow$ Nuclear Physics $\longrightarrow$ High
Energy Physics \\
\\ \hline
\end{tabular}
\end{center}

\noindent HEP is the front end or cutting edge of the human intellect
advancing into the unknown territory in its inward bound journey.
This {\it 100-year-long history} must be viewed together, to get a true
picture of HEP. It is within this broad framework that we must place any
particular contribution or the totality of Indian contributions, for a
proper perspective. Viewed in this light, it is perfectly natural to
include the great Indian contributions made in the earlier part of the
20th century. {\it If Bose, Raman \& Saha were alive and young today,
they would be doing HEP}. So, I start with their contributions \ldots

\smallskip
\begin{enumerate}
\item M N Saha's (1923) theory of thermal ionization played a crucial
role in the elucidation of stellar spectra and thus was of fundamental
importance for the progress of Astrophysics.
(Saha's (1936) reinterpretation of Dirac's quantization condition for
monopoles, in terms of ang. mom. quantization, was very original and its
importance is now recognized.)

\smallskip
\item S N Bose (1924) discovered Quantum Statistics even before the
discovery of Quantum Mechanics by Heisenberg \& Schr\"{o}dinger one year
later. Logically pursued, Bose's discovery by itself would have led to
Quantum Mechanics. But, History went differently. QM was discovered soon
(in fact, too soon) and the flood gates were open. This was unfortunate
for India. This may be called MISSED OPPORTUNITY I.

\smallskip
\item C V Raman (1928) discovered the inelastic scattering of photon on
bound electrons and thus took the concept of photon one step higher.
Raman effect is a fundamental experimental discovery that has not been
surpassed or even equalled in its importance and impact even after 70
years by any other experiment done in this country.

\smallskip
\item S.Chandrasekhar (1932) applied relativistic quantum mechanics to
the interior of stars. He calculated the degeneracy pressure (or Pauli
pressure) of a relativistic electron gas and thus initiated our
understanding of the gravitational collapse of stars.

\smallskip
\item H.J. Bhabha's (1935) calculation of $e^+e^-$ scattering was one of
the earliest nontrivial applications of Dirac equation to a process in
which Dirac's hole theory played a crucial role. This was done even
before a full-fledged Quantum Field Theory existed.
\end{enumerate}

\newpage

\section{HEP before and after the Standard Model}
\subsection{QED}
\begin{center}
\begin{tabular}{|c|} \hline \\
$L \ = \ - \frac{1}{4} (\partial_\mu A_\nu - \partial_\nu A_\mu)^2 \ -
\ \bar{\psi} \gamma^\mu (i \partial_{\mu} - e A_\mu) \psi - m \bar{\psi} \psi $
\\
\\ \hline
\end{tabular}
\end{center}

\begin{tabular}{lcl}
Planck (1900) & : & Quantization of radiation energy \\
[3mm]
Einstein (1905) & : & Photon \\
[3mm]
{\it Bose} (1924) & : & Photons as identical particles \\
[3mm]

$\left.\!\!\!\! \begin{tabular}{ll}{\it Bose} and Einstein & (1926)\\
  Fermi and Dirac & (1926) \end{tabular}\right\} $
  & : & Quantum Statistics\\[5mm]

Dirac (1927) & : & lays the {\it foundations for QED} by \\
& & introduction of $a,a^+$ for photons \\
& & \hfill $[a,a^+]=1$ (bosons) \\
[3mm]
Jordan \& Wigner (1928) & : &  $\{b,b^+\} \ = \ 1$ (fermions) \\
[3mm] $ \left.\!\!\!\! {\begin{tabular}{ll} {\rm Compton} &
(1925) \\
{\it Raman} & (1928)
\end{tabular}} \right\}$
& : & photon \ scatters \ like \ particle  \\
[3mm]
Dirac (1928) & : & Relativistic Eq. for electron \\
[3mm]
Anderson (1932) & : & Discovery of $e^+$ \\
[3mm]
{\it Bhabha} (1935) & : & $e^+ e^-$ scattering \\
[3mm]
Kramers (1947) & : & The idea of renormalization \\
[3mm]
Lamb \& Retherford (1947) & : & Exptl discovery of Lamb shift \\
Bethe (1947) & : & First calculation of Lamb shift \\
& & using renormalization
\end{tabular}

\begin{tabular}{llll}
$\left.\!\!\!\!\begin{tabular}{l} Feynman, \\ Schwinger, \\
Tomonoga, \\ Dyson \end{tabular} \right\}$ \  \  & $(1946-50)$ &:&
$\!\!\!\!\left\{ \begin{tabular}{l}Covariant Formalism, \\ Perturbation Series for
S matrix, \\ Feynman Diagrams, \\ QED emerges as a  \\
renormalizable QFT. \end{tabular} \right.$
\end{tabular}

The development of QED is concurrent with the development of QFT, which
has been the basic language of HEP, at least so far.
One may think of the history of QED originating with the concept of
the photon. But it was Dirac who laid the foundation for QED. The work of
Bose, Einstein, Dirac and Fermi led to the recognition that there are two
types of field quantization. \\

Modern QED starts with renormalization. By 1950, QED emerges as a
covariant, local and renormalizable QFT.
Here one must include S. N. Gupta's contribution.\\

\noindent {\it S N Gupta} (1950)
It was S N Gupta (1950) who first constructed a manifestly relativistic
formulation of QED. Before him, QED was formulated in coulomb gauge
which lacked manifest relativistic invariance. Gupta showed how to do
QED in the covariant Lorentz gauge in a consistent way using indefinite
metric.

\subsection{Weak Interactions}

\begin{tabular}{lcl}
Becquerel (1896) & : & Radioactivity $(\alpha, \beta, \gamma)$\\
\\
Pauli (1930) & : & "Neutrino" \\
\\
Fermi (1934) & : & Theory of $\beta$ - decay : $\displaystyle L_{int} \ = \ \frac{G_F}{\sqrt{2}} \bar{p} \gamma_{\mu} n \ \bar{e} \gamma^{\mu} \nu + h.c.$
\end{tabular}\\

\noindent
\begin{tabular}{lll}
$\!\!\!\! \begin{tabular}{ll}
\  Lee, Yang, Wu  & (1956)\ \ \ \ : \\
{\it Sudarshan} \& Marshak &(1957)\ \ \ \ : \end{tabular}$ & &
$\left.\!\!\!\!
\begin{tabular}{l}
Parity Revolution\\V-A Form
\end{tabular}   \right\}\gamma_{\mu} \rightarrow \gamma_{\mu} (1-\gamma_5) $
\\[5mm]
Feynman \& Gell-Mann  (1957)  \ \ \  \hspace{0.6mm}  : &&  Universal current $\times$ current theory
\end{tabular}

\begin{center}
\begin{tabular}{|lrcl|}
\hline &&&\\
\hspace{2cm}&$\displaystyle L_{int}$ & =& $\displaystyle \frac{G_F}{2\sqrt{2}} (J^+_\mu J^-_\mu + J^-_\mu J^+_\mu)$ \\
&$\displaystyle J^+_\mu$ & = &$ \displaystyle \frac{1}{2} \bar{u} \gamma_{\mu} (1-\gamma_5)d +
            \frac{1}{2} \bar{\nu} \gamma_\mu (1-\gamma_5) e + \ldots $\\
&$\displaystyle J^-_\mu$ & = &$ \displaystyle \frac{1}{2} \bar{d} \gamma_\mu (1-\gamma_5)u +
                     \frac{1}{2} \bar{e} \gamma_\mu (1-r_5) \nu + \ldots$\\
&&& \\ \hline
\end{tabular}\end{center}

The important steps are Pauli's suggestion of neutrino, Fermi's
construction of his famous theory (which was  \underline{the most
important step}),
  parity violation discovered by Lee, Yang and Wu and V-A form of
interaction proposed by Sudarshan and Marshak (another important Indian
contribution). The developments culminated in the universal current x
current form of Feynman and Gell-Mann.
Actually the final form was nothing but Fermi's theory, incorporating
$100\%$ parity violation. It is important to note that Fermi himself chose
only the vector form for the weak interaction, based on analogy to
electromagnetism. The discovery of maximal parity violation then required
an equal mixture of V and A. Therefore the simple form of the weak
interaction written down by Fermi purely on an intuitive basis in 1934
stood the ground for almost 40 years, until it was replaced by the SM.\\

\subsection {Strong Interactions}

\begin{tabular}{l}
Yukawa (1934)\\ [2mm]
Heisenberg:, Isospin Symmetry \\[2mm]
Powel, Occialini \ldots: Discovery of $\pi$ (1947) \\[2mm]
$\triangle$ resonance: Fermi (1952)  \\[2mm]
Chew-Low Theory : (1954) \\[2mm]
Discovery of Strangeness: Gell-Mann \& Nishijima (1955) \\[2mm]
Resonances (1957-65) \\[2mm]
S-matrix Theory (1957-62)\ :  G.F.Chew\\ [2mm]
$SU(3):  \left\{ \begin{tabular}{l} Sakata, Gell-Mann, Neeman (1961) \\[2mm]
Discovery of $\Omega^-$ (1964) \end{tabular} \right.$ \\[5mm]
Quarks: Gell-Mann, Zweig (1964) \\[2mm]
Current algebra, PCAC, Chiral sym (1965-70) \\[2mm]
Scaling in DIS and partons :  Bjorken, Feynman (1967)\\[2mm]
SLAC expts : Taylor, Friedman, Kendal (1967) \\[2mm]
Discovery of Asymptotic Freedom of NAGT \& Birth of QCD (1973)
\end{tabular}
\baselineskip12pt

\smallskip
\parindent6mm

Strong interactions proved a harder nut to crack. Going quickly over the
early history starting with Yukawa, let us come to the late fifties and
early sixties. Under the deluge of hundreds of hadrons (or hadronic resonances)
that were being discovered almost every week, QFT was declared dead and an
alternative philosophy called {\underline {S matrix theory}} was proposed, its
chief proponent being G. F. Chew. Many important ideas were developed under its
banner - dispersion relations, Regge poles, bootstrap, nuclear democracy
etc. Ultimately this proved to be a dead end. And a different line of
attack spearheaded by Gell-Mann proved more successful. Starting with
SU(3), this led to current algebra, and then to quarks, which finally led,
via scaling in DIS and asymptotic freedom to QCD. So, back to QFT even for
strong interactions.\\

However, one must not conclude that S matrix approach was a complete
failure. Although it was a failure for  hadrons, it is this approach that gave
birth to String Theory!\\

\noindent{\it The Bangalore event}

Here I want to describe an event that occurred in Bangalore in August, 1961. We
were having our first TIFR Summer School in Bangalore. Dalitz and
Gell-Mann were the lecturers. I was only a student, but people like
Bhabha, Menon and S. N. Biswas were also participants.Gell-Mann lectured on
SU(3) and the Eightfold Way-fresh from the anvil, even before they were
published. [This was flavor SU(3)-much before the days of color.] During
one of the lectures, Dalitz questioned him about the triplets. Why is he
ignoring them? Gell-Mann hedged and hawed in his characteristic fashion
and did not answer the question inspite of Dalitz's repeated questions. If
Gell-Mann had tried to answer the question, quarks would have been born in
Bangalore in 1961 instead of having to wait for another three years. If
any of the other participants had succeeded answering, we would have
got the quarks and this would have been a major Indian contribution. So, I
would call this {\bf Missed Opportunity II.}\\

\subsection{Summary of HEP before the Standard Model (before circa 1971)}

Putting together(2.1)-(2.3), we have

\begin{eqnarray*}
L & = & - \frac{1}{4} F_{\mu \nu} F^{\mu \nu} + \bar{e} \left[ i\gamma_\mu
(\partial^\mu - ie A^\mu)-m_e\right] e + i \bar{\nu}_e \gamma_\mu
\partial^\mu \nu_e \\
&&+ \bar{\mu} \left[i \gamma_\lambda(\partial^\lambda - ie
A^\lambda)-m_\mu\right]\mu + i \bar{\nu}_\mu \gamma_\lambda \partial^\lambda
\nu_\mu\\
&& + \bar{u} \left[ i\gamma_\mu(\partial^\mu + \frac{2}{3} i e A^\mu) -
m_u\right]u \ + \ \bar{d} \left[i\gamma_\mu(\partial^\mu - \frac{i}{3}
eA^\mu) - m_d\right]d\\
&& + \bar{s} \left[ir_\mu(\partial^\mu - \frac{1}{3} i e A^\mu) -
m_s\right]s \\
&& + \frac{G_F}{2\sqrt{2}} \left(J_\mu^+ J^-_\mu + J^-_\mu J^+_\mu\right)\\
&& +\mbox{\ strong interactions among quarks whose nature was not known}.
\end{eqnarray*}
where,

$$ J_{\lambda}^{-}={1 \over 2} \bar{e} \gamma_{\lambda}(1 -
\gamma_{5}) \nu_{e} + {1 \over 2} \bar{\mu} \gamma_{\lambda}(1 -
\gamma_{5}) \nu_{\mu}
  + {1 \over 2} (\bar{d} \cos \theta_{c} + \bar{s} \sin \theta_{c})
\gamma_{\lambda} (1 - \gamma_{5}) u $$

$$J_{\lambda}^{+}= (J_{\lambda}^{-})^\dagger $$
and
$$ \sin \theta_{c} \approx 0.22. $$

\noindent
Here we have  the Lagrangian density describing the em and weak interactions of
the three quarks $u, d, s$ and the four leptons $e, \mu, \nu_e,\nu_\mu$. The
existence of these quarks as the constituents of the hadrons had been
inferred from hadron spectroscopy through a clever guess. However nobody
knew the form of the strong interaction among the quarks which is
responsible for binding them into hadrons. So, it is left unspecified.
The weak current $ J_\lambda^-$ has been written in term of the Cabibbo-rotated
quarks, in order to incorporate the weak decays of the strange hadrons. CP violation
was experimentally known, but \underline{not} understood theoretically.\\

\subsection{ The Standard Model of High Energy Physics }

\begin{tabular}{llll}
&{\bf Theory} & & {\bf Experiment}\\
1954 & Nonabelian gauge fields &&\\
1964 & Higgs mechanism & & \\
1967 & EW Theory & & \\
& &1968 & Scaling in DIS \\
1971 & Renormalizability of EW Theory \hskip .5cm & & \\
1973 & Asymptotic freedom $~~\rightarrow$ QCD & 1973 & Neutral current \\
& & 1974 & Charm \\
& & 1975 & $\tau$-lepton \\
& & 1977 & Beauty \\
& & 1978 & polarized $e\, d$ expt \\
& & 1979 & 3 jets \\
& & 1983 & W,Z Bosons\\
& & 1994 & Top \\
& & 1998 & $\nu$ mass(?)
\end{tabular}

\vspace{3mm}

The major events which culminated in the construction of the Standard Model
are shown in this table in chronological order. Using nonabelian gauge
theory with Higgs mechanism, the EW theory was already constructed in 1967,
although it attracted the attention of most theorists only after another 4
years, when it was shown to be renormalizable. The discovery of asymptotic freedom of NAGT
and the birth of QCD in 1973 were the final inputs that led to the full
standard model.\\

On the experimental side, the discovery of scaling in DIS which led to the
asymptotically free QCD and the discovery of the NC which helped to confirm
the EW theory can be regarded as crucial experiments. To this list, one may add the
polarized electron deuteron experiment which showed that SU(2) x U(1) is the correct group for EW
theory, the discovery of gluonic jets in $ e^{+} e^{-} $ annihilation confirming
QCD and the discovery of W and Z in 1983 that established the EW theory. The
experimental discoveries of charm, $\tau$, beauty and top were also
fundamental for the concrete 3-generation SM.\\

However note the blank after 1973 on the theoretical side. Theoretical
physicists have been working even after 1973 and experiments also are
being done. But the tragic fact
is that none of the bright ideas proposed by theorists in the past 27
years has received any experimental support. On the other side,
experiments  have
only been confirming the theoretical structure completed in 1973.
None of the experiments done since 1975 has made any independent
discovery (except the discovery of neutrino mass).  If this
continues for long, it will be too bad for the future of HEP. I shall come
back to this point later.

\vspace{3mm}
The complete Lagrangian of the Standard Model is given by
\newpage
\begin{eqnarray*}
  {\cal L} & =& -{1 \over 4}(\partial_{\mu}G_{\nu}^{i} - \partial_{\nu}G_{\mu}^{i} - g_{3}
     f^{ijk}G_{\mu}^{j} G_{\nu}^{k})^2  \\
       &&  -{1 \over 4}(\partial_{\mu}W_{\nu}^{a} - \partial_{\nu}W_{\mu}^{a} - g_{2}   \epsilon^{abc}W_{\mu}^{b} W_{\nu}^{c})^2 - {1 \over 4} (\partial_{\mu} B_{\nu}
          - \partial_{\nu} B_{\mu})^2  \\
 	    && -\sum_{n} \bar{q}_{nL} \gamma^\mu (\partial_{\mu} + ig_{3} {\lambda^{i}
 	    \over 2} G_{\mu}^{i} +ig_2 {\tau^a \over 2} W_\mu^a +{ig_1 \over 6}
 	    B_{\mu}) q_{nL}  \\
 	    &&   -\sum_{n} \bar{u}_{nR} \gamma^\mu (\partial_{\mu} + ig_{3} {\lambda^{i}
 	    \over 2} G_{\mu}^{i} + i{2 \over 3} g_1 B_\mu) u_{nR}  \\
 	    &&   -\sum_{n} \bar{d}_{nR} \gamma^\mu (\partial_{\mu} + ig_{3} {\lambda^{i}
 	    \over 2} G_{\mu}^{i} - i{g_1 \over 3} B_\mu) d_{nR}  \\
 	    && -\sum_{n} \bar{l}_{nL} \gamma^\mu (\partial_{\mu} + ig_{2} {\tau^{a}
 	    \over 2} W_{\mu}^{a} - i{g_1 \over 2} B_\mu) l_{nl}  \\
 	    &&   -\sum_{n} \bar{e}_{nR} \gamma^\mu (\partial_{\mu} - ig_{1} B_\mu)
 	    e_{nR}  \\
 	    && + | (\partial_{\mu} + ig_{2} {\tau^{a} \over 2} W_{\mu}^{a} - i{g_1 \over 2}B_\mu) \phi |^2 - \lambda(\phi^+ \phi - v^2)^2  \\
 	    && - \sum_{m,n} (\Gamma_{mn}^u \bar{q}_{mL} \phi^c u_{nR} + \Gamma_{mn}^d
 	    \bar{q}_{mL} \phi d_{nR} + \Gamma_{mn}^e \bar{l}_{mL} \phi e_{nR} + h.c.)
 	    \\
 	    \end{eqnarray*}

STANDARD MODEL is the basis of {\it all} that is known
in HEP. Although it is believed that SM is only an effective
low energy description and it is to be replaced by something
beyond, so far SM has resisted all attempts at overthrowing
it. All the precision tests performed so far are in beautiful
agreement with SM. All the experimental signals that seem
to signal its overthrow, disappear in about 6 months -- 1
year, except one signal, namely the signal that {\it
NEUTRINO} has mass.
Neutrino is the only particle, a part of which (its RH
part) has zero quantum number and so it is not acted on by
the SM group : $SU(3) \times SU(2) \times U(1)$. So, its RH
part is absent in SM \& as a consequence, the nautural mass
of neutrino within SM is zero. That is why $\nu$ having a
mass is regarded as a signal beyond SM.

Note the almost complete absence of Indian contribution. (Of course Salam's
name is there, as a major contributor to the construction of the $ SU(2)
\times U(1) $ electroweak theory. We shall eschew parochialism and include
him since he is from the subcontinent.) Let me give  a little bit of my
side of the story here. I was aware of the beauty of Yang-Mills (YM) theory from the
time of Sakurai's famous Annals of Physics paper of 1960 and I realized
the importance of YM theory to weak interaction ever since I listened to
Veltman in the Varenna School in 1964 where he stressed the
conservation of weak currents. When Weinberg's paper with the quaint title
"A model of Leptons" came out in Physical Review Letters in 1967, I was
immediately convinced that this was the correct theory for weak
interactions and began to work on it. I still missed the boat completely
because I was too muddle-headed and stupid. Instead of trying to
renormalize the divergences away (which we now know to be the right thing,
after t'Hooft showed it in 1971), I was trying to generate the strong
interactions from the divergences. I was too ambitious and missed the
real thing.\\

Mine was a double failure. Since I was very familiar with partons, scaling
and the quark-model sum rules that the DIS structure functions were found
to obey, I was fully aware of the serious problem that was staring at
everybody's face, namely, how to reconcile the free-quark behavior
exhibited by the DIS structure functions with the superstrong interactions
of quarks inside the hadrons. The techniques that were subsequently used to
effect the reconciliation were also known to me. In fact I was giving
a series of lectures on Wilson's RG ideas and the Callan-Symanzik $
\beta $ function at TIFR, when the preprints of Politzer and
Gross-Wilczek proving asymptolic freedom in YM theory came out.\\

So, I failed on both fronts :  on both the two important field-theoretic
discoveries of the latter part of the $ 20^{th} $ century - namely
renormalizability of YM theory with SSB and asymptotic freedom of YM
theory - both of which being the essential theoretical inputs in the
construction of the SM of HEP. This is MISSED OPPORTUNITY III.\\

Forgetting about myself, it was a collective failure of the Indian High
Energy Physicists. By that tme we had strong theory groups in the country
and we should have made significant contributions in the construction of
SM, but we did not. In my opinon, this is a glaring failure and we cannot
  forgive ourselves for it.\\

\section{Future of HEP}

Standard Model is not the end of the story. There are too many
loopholes in  it. First of all, there are many interesting
questions and unsolved problems within SM :
\begin{itemize}
\item  Higgs and symmetry breaking
\item QCD and Confinement
\item Neutrinos and  their masses and mixings
\item CP and its violation
\end{itemize}

The solution of these problems may already take us {\it beyond
SM}.

However, the biggest loophole in SM is the omission of
gravitation, the most important force of nature. Hence, it is
now recognized that {\it quantum gravity} (QG) is the next frontier
of HEP, and that {\it the true fundamental scale of physics
is the Planck energy} $10^{19}$ Gev, which is the scale of
QG.

We are now probing the region with energy $\le 10^3 GeV $. One can
see the vastness of the domain one has to cover before QG is
incorporated into physics. In their attempts to probe this
domain of $10^{3} - 10^{19}$ Gev, theoretical physicists have
invented many ideas such as SUSY, hidden dimensions etc and
based on these ideas, they have constructed many beautiful
theories, the best among them being the string theory, which
may turn out to be the correct theory of QG.

Especially after the second string revolution of 94-96 that led
to breakthroughs such as Duality linking many string
theories, multidimensional branes \ldots   String Theory
(~M~Theory~?) has become very rich. But, Physics is not theory
alone. Even beautiful theories have to be confronted with
experiments and either confirmed or thrown out. Here we
encounter a serious crisis facing HEP. In the next 10-25
years, new accelerator facilities with higher energies such
as the LHC ($\sim 10^4$ Gev) or the Linear Electron Collider
will be built so that the prospects for HEP in the immediate
future appear to be bright. Beyond that period, the
accelerator route seems to be closed because known
acceleration methods cannot take us beyond about $10^5$ GeV.
It is here that one turns to hints of new physics from {\it
Cosmology, Astroparticle Physics and Nonaccelerator particle
physics}. However, these must be regarded as only our first and
preliminary attack on the unknown frontier. These are only
hints. Physicists cannot remain satisfied with hints and
indirect attacks on the superhigh energy frontier.

\noindent {\bf To sum up the situation :}
There are many interesting fundamental theories taking us to
the Planck scale and even beyond, but unless the experimental
barrier is crossed, these will remain only as Metaphysical
Theories.  It follows that,

\begin{itemize}
\item {\it either,  new ideas of acceleration have to be
discovered},
\item {\it or, there will be an end to HEP by  about 2020
AD.}
\end{itemize}

Some of the ideas being pursued are laser beat-wave method,
plasma wake field accelerator, laser-driven grating linac,
inverse free electron laser, inverse Cerenkov acceleration
etc. What we need, are a hundred crazy ideas. May be, one of
them will work!

By an optimistic extrapolation of the growth of accelerator
technology in the past 60 years, one can show that {\it
$10^{19}$ GeV can be reached in the year 2086}. (See my
Calcutta talks)\footnote{\noindent
  Perspectives in High Energy Physics, Proceedings of the VIII HEP
  Symposium, Calcutta (1986), p 399;\\
  The Future of HEP, Particle Phenomenology in the 90's, World Scientific (1992), p1
  } But, this is possible only if newer methods and newer
technologies are continuously invented.\\

\noindent {\bf Another Way Out}

In the past three years, another revolutionary idea is
being-tried -- namely to bring down Planck scale from $10^{19}$
GeV to $10^3$ GeV.  This is the so called TeV scale gravity
which uses large (sub-mm) extra dimensions.  (If we cannot go up
to the mountain top why not ask the mountain
top to come down?)

This is a very interesting field, with a bewildering variety of
worlds that theorists can construct, as a scan of recent hep-net
will show.

Is Nature so kind and considerate to us, that it would have
brought down the Planck scale for our sake? Only Future can tell.

But, if this turns out to be correct, then Quantum Gravity and
String Theory are not some distant theories relevant at $10^{19}$
GeV, but they are immediately relevant at $10^3 - 10^5$ GeV. So,
it becomes even more urgent to understand String Theories and
assimilate them into Physics!

\section { Status of HEP in India and  Suggestions for the Future}

\noindent {\bf Theory :}
There is extensive activity in HEP theory in the country, spread
over TIFR, PRL, IMSc, SINP, IOP, MRI, IISc, Delhi University,
Panjab University, BHU, NEHU, Gauhati University, Hyderabad
University, Cochin Univesity, Viswabharati, Calcutta University,
Jadavpur University, Rajasthan University and a few other
centres. Research is done in almost all the areas in the field,
as any survey will indicate.
Theoretical HEP continues to attract the best students and as a
consequence its future in the country appears bright. However, it
must be mentioned that this important national resource is being
underutilized. Well-trained HEP theorists are ideally suited to
teach any of the basic components of Physics such as QM, Relativity,
QFT, Gravitation and Cosmology, Many Body Theory,  Statistical Mechanics,
and Advanced Mathematical Physics
since all these ingredients go to make up the present-day HEP
Theory. Right now, most of these bright young theoretical
physicists are seeking placement in the Research Institutions.
Ways must be found so that a larger fraction of them can be
absorbed in the Universities. Even if just one of them joins each
of the 200 Universities in the country, there will be a
qualitative improvement in physics teaching throughout the
country. But, this will not happen unless the young theoreticians
gain a broad perspective and train themselves for teaching-cum
research careers. Simultaneously, the electronic communication
facilities linking the Universities among themselves and with the
Research Institutions must improve. This will solve the
frustrating isolation problem which all the University
Departments face.

\noindent{\bf Experiments:}

Many Indian groups from National Laboratories as well as
Universities (TIFR, VECC, IOP, Delhi, Panjab, Jammu and Rajasthan
Universities) have been participating in 3 major international
collaboration expts :

\begin{itemize}
\item $L3$ expt on $e^+e^-$ collisions at LEP (CERN)
\item $D\ zero$ expt on $\bar{p}p$ collisions at the Tevatron
(Fermilab)
\item WA93 \& WA 98 expts on heavy-ion collisions at CERN

\end{itemize}

As a result of the above experience, the Indian groups are well
poised to take advantage of the next generation of colliders such
as Linear Collider and LHC. Already the Indian groups have joined the
international collaboration in charge of the CMS which will be
one of the 2 detectors at LHC. It is also appropriate to mention
here that Indian engineers and physicists are contributing
towards the construction of LHC itself.

Thus, the only experimental program that is pursued in the
country is the participation of Indian groups in international
accelerator-based experiments. This is inevitable at the present
stage, because of the nature of present-day HEP experiments that
involve accelerators, detectors, experimental groups and
financial resources that are all gigantic in magnitude.

While our participation in international collaborations must
continue with full vigor, at the same time, for a balanced
growth of experimental HEP, we must have in-house activities
also. Construction of an accelerator in India, in a suitable
energy range which may be initially 10-20 GeV and its utilization
for research as well as student-training will provide this
missing link.

In view of the importance of underground laboratories in $\nu$
physics, monopole search, $p$ decay etc, the closure of the deep
mines at KGF is a serious loss. This must be at least partially
made up by the identification of some suitable site and we must
develop it as an underground lab for nonaccelerator particle
physics.\footnote{ A proposal for the construction of an Indian Neutrino
Observatory (INO) is under preparation. Hopefully this project will be
undertaken by a collaboration between various institutions in the country}

Finally, it is becoming increasingly clear that known methods of
acceleration cannot take us beyond tens of TeV. Hence in order to
ensure the continuing vigor of HEP in the 21st century, it is
absolutely essential to discover new principles of acceleration.
Here lies an opportunity that our country should not miss! I have
been repeatedly emphasizing for the past many years $(> 10)$ that
we must form a small group of young people whose mission shall be
to discover new methods of acceleration.\footnote{IPR is already initiating
research in this area and CAT is training young scientists in accelerator
technology through SERC Schools }

To sum up, a 4-way program for the future of experimental HEP in
this country is suggested.

\begin{enumerate}
\item A vigorous participation of Indian groups in international
experiments, accelerator-based as well as nonaccelerator-based.
\item Construction of an accelerator in this country.
\item Identification and development of a suitable underground lab
for nonaccelerator particle physics, especially neutrino physics.
\item A programme for the search for new methods of acceleration
that can take HEP beyond the TeV energies.
\end{enumerate}

\newpage
\section{A list of more Indian contributions}

The following list is presented with apologies to the missing
names.

\begin{center}
  {\bf Theory} \hspace{60mm}  {\bf Experiment} \hfill \\[3mm]
\begin{tabular}{l|l}
\hline\\[2mm]
  \parbox[t]{70mm} { Bhabha and Heitler: Theory of Cosmic\\\hspace*{\fill} ray showers} &
  \parbox[t]{70mm} {Menon \& others: Discovery of $K$ decay \\ \hspace*{\fill} modes }  \\[2mm]
\parbox[t]{70mm} { Udgaonkar: Regge asymptotics for\\ \hspace*{\fill} hadron  cross sections}&
    \parbox[t]{70mm} { TIFR group : Pioneered \\ \hspace*{\fill} many cosmic ray expts } \\ &  \\
\parbox[t]{70mm}{ S M Roy : Integral Eq. for $\pi \pi$}&
  \parbox[t]{70mm} { Kolar: First detection of atmospheric $\nu$, \\ \hspace*{\fill}  Search for $p$ decay}     \\[4mm]\parbox[t]{70mm}{ V~Gupta and  V~Singh : $SU(3)$ sum rule \\ \hspace*{\fill} for widths of baryon decuplet}&
       \parbox[t]{70mm} { Baba, Indumathi etal: Axion~search } \\ &\\
\parbox[t]{70mm}{Beg \& V Singh : SU(6) mass formula } &
  \parbox[t]{70mm} { Cowsik, Krishnan, Unnikrishnan : Search for fifth force }  \\[2mm]
\parbox[t]{70mm}{Mathur, Pandit \& others : Soft $\pi$ \\ \hspace*{\fill} relation for $K_{\ell_3}$ form factor }&
   \\ &\\
\parbox[t]{70mm}{Tapas Das, Mathur \& others : $\pi^+\pi^o$ \\ \hspace*{\fill} mass  diff from current algebra }&
   \\ & \\
  \parbox[t]{70mm}{Balachandran \& others : $\pi N$ \\ \hspace*{\fill} scattering length from current algebra }&
\\
\parbox[t]{70mm} { Cowsik \& others : Bound on $m_\nu$  from \\ \hspace*{\fill} Cosmology}  &
$\left.{ \begin{tabular}{l}Precision tests of SM \\ Discovery of Top \\Search for QGP \end{tabular} }\right\}$ \begin{tabular}{l}Indian~groups\\have partici-\\pated in these\\ experiments\end{tabular} \\
\hline
\end{tabular}

\vspace{4mm}
\parbox[t]{120mm}{ These horizontal lines demarcate the boundary between established Physics and Physics that may be established in the $21^{st}$ century or later.}
\vspace{4mm}

\begin{tabular}{l|l}
\hline\\
\parbox[t]{70mm}{Pati \& Salam : GUT and p decay} &
\parbox[t]{70mm}{Pakvasa \& Raghavan : Testing for $\nu$ \\ \hspace*{\fill} oscillation through NC expts}\\[2mm]
\parbox[t]{70mm}{Mohapatra etal : See-saw, spontaneous \\ \hspace*{\fill} L violation} &
  \parbox[t]{70mm}{Raghavan : Borexino} \\&\\
\parbox[t]{70mm}{Kaul \& Majumdar : SUSY to solve \\ \hspace*{\fill} hierarchy} & \\&\\
  \parbox[t]{70mm}{Asoke Sen : Strings, pioneered Duality} & \\[2mm]
  \parbox[t]{70mm}{Ashtekhar : Quantum Gravity} &\\[2mm]
\hline
\end{tabular}
\end{center}


Inspite of the above list one must admit the remarkable absence of significant
Indian contributions in the recent history of HEP. Why have we come down? Where
are the equivalents of Bose and Raman in present-day HEP ?
There may be sociological reasons for this, but this is not the place to go into them.

Is it possible that
India throws up great names only when physics goes through revolutionary
development as in the beginning of the 20th century? If so, the next revolution
which may come in the 21st Century must be eagerly watched! Remember that the
solution of the Quantum Gravity problem and/or the formulation of String Theory is
still incomplete. That may usher in the next revolution  in Physics and may involve
  great contributions from India. There are already signs of this, in the quality of
  Indian contributions to String Theory and Quantum Gravity.  I will close with this
  optimistic remark.

\end{document}